\documentclass[apj]{emulateapj}
\usepackage{graphics}
\usepackage{subfigure}
\usepackage[caption=false]{subfig}
\usepackage{epsfig}
\usepackage{alltt}
\usepackage{verbatim}
\usepackage{amsmath}

\def\spose#1{\hbox to 0pt{#1\hss}}
\def\simlt{\mathrel{\spose{\lower 3pt\hbox{$\mathchar"218$}}
    \raise 2.0pt\hbox{$\mathchar"13C$}}}
\def\simgt{\mathrel{\spose{\lower 3pt\hbox{$\mathchar"218$}}
    \raise 2.0pt\hbox{$\mathchar"13E$}}}

\slugcomment{ApJ accepted}
\shortauthors{Comerford \& Simon}
\shorttitle{Preferential Accretion onto the Secondary Black Hole Strengthens Gravitational Wave Signals}

\begin{document}

\title{Preferential Accretion onto the Secondary Black Hole \\ Strengthens Gravitational Wave Signals}

\author{Julia M. Comerford\altaffilmark{1} and Joseph Simon\altaffilmark{1,2}}

\affil{$^1$Department of Astrophysical and Planetary Sciences, University of Colorado, Boulder, CO 80309, USA}
\affil{$^2$Department of Physics, Oregon State University, Corvallis, OR 97331, USA}

\begin{abstract}

Pulsar timing arrays have recently found evidence for nanohertz gravitational waves that are consistent with being produced by a cosmological population of binary supermassive black holes (SMBHs). However, the amplitude of this gravitational wave background is larger than predicted from theoretical and empirical models of SMBH binary populations. We investigate preferential accretion onto the secondary, less massive SMBH of the binary as a potential solution to this discrepancy. We carry out the first observationally-based analysis of the effect of preferential accretion on the SMBH binary population, and we find that preferential accretion onto the secondary SMBH increases the binary SMBH mass ratio, causing many minor galaxy mergers to lead to major SMBH mergers. The fraction of SMBH mergers that are major mergers increases by a factor of 2-3 when preferential accretion is included. Further, we find that only a small amount of preferential accretion (10\% total SMBH mass growth) is needed to bring the predicted gravitational wave background amplitude into agreement with observations. Preferential accretion has an even larger effect on gravitational wave signals detected by LISA, which will probe SMBH binaries at higher redshifts where the environment is more gas-rich, and can also help explain the rapid build up of overmassive black holes at high redshifts observed by the James Webb Space Telescope. It also shortens the time to the first detection of an individual SMBH binary emitting continuous waves. Preferential accretion strengthens the gravitational wave signals produced by any binary embedded in a circumbinary disk, including LIGO sources. 

\end{abstract}  

\section{Introduction}
 \label{intro}

Mergers of galaxies, each with its own central supermassive black hole (SMBH), are expected to create gravitationally-bound binary SMBH systems \citep{BE80.1}. When these binary SMBHs coalesce, they become the highest signal-to-noise ratio sources of gravitational waves in the universe. Pulsar timing arrays are tuned to the nanohertz gravitational waves produced by these binary SMBHs at $\simlt$milliparsec (mpc) separations. A gravitational wave background (GWB) is expected to be produced by the integrated emission in the nanohertz band from the full population of binary SMBHs over cosmic time (e.g., \citealt{PR72.1,SE04.1,BU19.1}).

All major pulsar timing array collaborations have recently reported evidence for a common spectrum, correlated noise process that is likely to be the GWB produced by a cosmological population of merging SMBH binaries (e.g., \citealt{AG23.1, EP23.1, RE23.1, XU23.1}). However, the observed GWB signal has a higher amplitude than expected from existing cosmological models of SMBH binaries (e.g., \citealt{ZH19.1, MI21.1, IZ22.1, AG23.2}). This could be explained by exotic sources such as cosmic inflation and cosmic strings (e.g., \citealt{EL21.1,VA21.1,AF23.1}). The mismatch between the expected and observed amplitude could also be due to the limitations of our understanding of SMBH formation, growth, and evolution across cosmic time, and several new studies have re-examined the assumptions that go into models of SMBH binaries in an effort to explain the discrepancy.

Several different parameters that SMBH binary population models are based on have been tested so far. For example, the higher GWB amplitude might be explained by variations in the way the SMBH mass is inferred. However, changes in the way the SMBH mass is estimated for models of the GWB have so far not been able to account for the discrepancy. This includes increasing the scatter in the $M_\bullet - \sigma$ relation used to derive SMBH mass from the galaxy stellar velocity dispersion, and systematic differences between using the relation between SMBH mass and stellar bulge mass $M_\bullet-M_{bulge}$ or $M_\bullet - \sigma$ to derive SMBH mass \citep{CA22.1,MA23.1,SA24.1}. 

If there are a larger number of $\simlt 10^{10} \; M_\odot$ SMBHs in the universe \citep{SA25.1}, or if SMBHs build up most of their mass at $z>1$ \citep{SO25.1}, then the GWB amplitude can increase. A combination of an increased number of mergers and an enhanced accretion rate of material onto the SMBHs can also increase the GWB amplitude \citep{SA25.2}. Additionally, a more detailed accounting of the largest SMBHs in the local Universe ($z \sim 0$) can help to alleviate some tensions \citep{LI24.1}.

Another unexplored possibility is that the mass ratios of the SMBH binaries are larger than expected (i.e., the two SMBHs are closer to equal mass), which would increase the amplitude of the GWB. The amplitude of the GWB depends on four key parameters -- the chirp mass of each source, the frequency of the gravitational waves emitted by each source, the redshift of each source, and the number density of binary SMBHs. The greatest dependence is on the chirp mass, which is defined as
\begin{equation}
    \mathcal{M} = \left[ \frac{q_\bullet}{(1 + q_\bullet)^2} \right]^{3/5} M_{\bullet, tot} \; ,
    \label{chirp}
\end{equation}
where $q_\bullet=M_{\bullet, 2}/M_{\bullet, 1}$ is the ratio of the mass of the secondary (less massive) SMBH $M_{\bullet, 2}$ to the mass of the primary (more massive) SMBH $M_{\bullet, 1}$ and the total SMBH mass is $M_{\bullet, tot}=M_{\bullet, 1}+M_{\bullet, 2}$. The amplitude of the GWB depends on $\mathcal{M}^{5/6}$ \citep{PH01.1}, so that any evolution in $q_\bullet$ would have a strong influence on the amplitude.

Current models of binary SMBHs assume that the SMBH mass ratio $q_\bullet$ does not change as the binaries evolve through a galaxy merger (e.g., \citealt{SE13.1,IZ24.1}). However, observations of active galactic nuclei (AGNs) in galaxy mergers show that SMBHs grow in mass as they progress through a galaxy merger (e.g., \citealt{EL11.1,KO12.1,FU18.1,ST21.1,CO24.1}). More importantly, observational measurements of the Eddington ratios of AGNs in galaxy mergers show preferential mass growth in the SMBH in the less massive galaxy (e.g., \citealt{LI11.3,CO15.1,BA23.1}). Hydrodynamical simulations of SMBHs in merging galaxies show the same effect \citep{CA15.1}. Such preferential mass growth in the secondary SMBH is important to explore in the context of gravitational waves, as it would drive increases in $q_\bullet$ and therefore in the GWB amplitude \citep{SI20.1}.

In this paper, we incorporate preferential accretion onto the secondary SMBH into an observationally-based model of the GWB for the first time. We build on the framework from \citet{SI23.1}, using the most up-to-date observational inputs. We focus on modeling massive galaxies (stellar masses $>10^{10} \, M_\odot$) out to $z=1.5$, as these are the galaxies that produce the binary SMBH population that appreciably contributes to the GWB (e.g., \citealt{SE13.1,RA15.1,SI16.1}). As in previous work, we use observational inputs whenever available to build a population of binary SMBHs that produce a predicted GWB that we can compare to pulsar timing array observations.
One notable update is that we use a self-consistent, observationally-derived galaxy merger rate, which is based on  spectroscopic close galaxy pairs in the Sloan Digital Sky Survey (SDSS). 

This paper is organized as follows. In Section~\ref{analysis}, we present the inputs that we use to derive a SMBH binary population. In Section~\ref{results}, we show the impact of differential accretion on the fraction of SMBH mergers that are major mergers; translate the SMBH binary population into a GWB prediction; show the impact of preferential mass growth in the secondary SMBH on the amplitude of the GWB; and present the implications of differential accretion for continuous wave sources, LISA massive black hole binaries, JWST high redshift black holes, and LIGO stellar mass black hole binaries. Finally, Section~\ref{conclusions} summarizes our conclusions.

Throughout this work, we assume a Hubble constant $H_0 =70$ km s$^{-1}$ Mpc$^{-1}$, $\Omega_m=0.3$, and $\Omega_\Lambda=0.7$.

\section{Analysis}
\label{analysis}

Here we present the variables that we use to model the cosmological population of SMBH binaries and compute a predicted GWB amplitude. We use observationally-based inputs wherever they are available.

\subsection{Galaxy Merger Rate}
\label{mergerrate}

To model the population of binary SMBHs, we begin with the population of merging galaxies. We start with a galaxy merger rate that is based on close spectroscopic pairs of galaxies in SDSS \citep{SI25.1}. The parent population is a $M_* \simgt 10^9 \, M_\odot$ mass-complete sample of SDSS DR7 galaxies at $0.02 \le z \le 0.2$. The spectroscopic galaxy pairs are defined by line-of-sight velocity separations $\Delta v < 500$ km s$^{-1}$, projected spatial separations $5 \le r_p (\rm kpc) \le 100$, and galaxy stellar mass ratios $0.1 \le q_{gal} \le 1$. 

Using this sample of close pairs, the $0.02 \le z \le 0.2$ galaxy merger rate is
\begin{equation}
    \mathcal{R}_{gal}(q_{gal}, M_*) =\frac{C_{merg}(M_*, r_p, \Delta v) \; \; f_{pair}(q_{gal}, M_*)
    }{<T_{obs}(M_*, r_p, \Delta v)>} \, ,
\end{equation}
where $C_{merg}$ is the correction factor to translate the number of galaxy pairs to the number of actual galaxy mergers (measured from close pairs of galaxies in the empirical model {\tt EMERGE}; \citealt{OL21.1}), $f_{pair}$ is the fraction of galaxies that are close pairs (including a factor to account for small angle incompleteness in SDSS), and $<T_{obs}>$ is the cosmologically averaged merger timescale (measured again from close pairs of galaxies in {\tt EMERGE}; \citealt{OL21.1}).

The $0.02 \le z \le 0.2$ galaxy merger rate we use here is
\begin{equation}
\mathcal{R}_{gal}(q_{gal}, M_*> 10^{10} M_\odot) = \frac{0.04}{q_{gal}} \; \mathrm{Gyr}^{-1} \; ,
\end{equation}
where $M_* > 10^{10}$ $M_\odot$ is the galaxy stellar mass limit of the sample.

This galaxy merger rate calculation has several major updates as compared to previous measurements that also used spectroscopic close pairs in SDSS (e.g., \citealt{LO11.1}). One update is that it incorporates $C_{merg}$ and $<T_{obs}>$ formulas taken from the same suite of simulations ({\tt EMERGE}; \citealt{OL21.1}), rather than combining disparate results together. Additionally, both formulas are more complex than those used in previous studies. For instance, instead of using only a single value with a modest dependence on $M_*$ for $<T_{obs}>$ \citep{SI16.1}, both formulas are functions of $r_p$, $\Delta v$, and $M_*$. 

Since the galaxy merger rate was only derived for $0.02 \le z \le 0.2$ galaxies, due to the redshift limitations of SDSS, it does not provide the means to measure the evolution of the galaxy merger rate with redshift. For the galaxy merger rate used in this work, we apply a redshift scaling of $(1+z)^{1.8 \pm 0.2}$, which is a combination of $f_{pair} \sim (1 + z)^{0.8 \pm 0.2}$ observed by \citet{DU19.1} and $<T_{obs}> \sim (1 + z)^{-1}$ from the empirical model {\tt EMERGE} \citep{OL21.1}. 

\subsection{SMBH Mass}
\label{bhmass}

To estimate SMBH masses for any large population of galaxies outside of the local universe, such as the SDSS galaxy population here, we must rely on scaling relations between SMBH mass and host galaxy properties. The galaxy stellar bulge mass and stellar bulge velocity dispersion are two common ways to infer the SMBH mass $M_\bullet$, via the $M_\bullet-M_{bulge}$ and $M_\bullet - \sigma$ scaling relations, respectively (e.g., \citealt{FE00.1, GE00.1, HA04.1, GU09.2}). Studies have suggested that $M_\bullet - \sigma$ is the more fundamental relation (e.g., \citealt{BE12.2,MA20.1,HU25.1,SH25.1}), while $M_\bullet-M_{bulge}$ may simply be a projection of $M_\bullet - \sigma$ (e.g., \citealt{VA16.1}). Here, we use the $M_\bullet - \sigma$ relation to derive SMBH masses.

We begin with a velocity dispersion function based on spectroscopic measurements of velocity dispersion where possible, and extended to higher redshifts using inferred velocity dispersions. For the spectroscopic measurements of velocity dispersion, we use SDSS data for $z<0.3$ \citep{BE10.2} and the Large Early Galaxy Astrophysics Census (LEGA-C) data for $0.5 < z < 1$ \citep{TA22.1}. Due to the lack of spectroscopic surveys of galaxies at higher redshifts, we use inferred velocity dispersions to fill out the velocity dispersion function at $1 < z < 1.5$. Inferred velocity dispersions are derived from the virial theorem and the observations that a galaxy's stellar mass is proportional to its dynamical mass (e.g., \citealt{TA10.1}). This inferred velocity dispersion has been shown to agree well with spectroscopic measurements of velocity dispersion (e.g., \citealt{BE11.2,TA22.1,HU25.1}). We use the inferred velocity dispersion function from \cite{BE12.3} for velocity dispersions at $1 < z < 1.5$. Our end result is a velocity dispersion function that is based on spectroscopic velocity dispersions at $z<1$ and inferred velocity dispersions at $1 < z < 1.5$.

Then, we derive SMBH masses from the velocity dispersion function, using the $M_\bullet - \sigma$ relation of \cite{DE19.1}. The \cite{DE19.1} $M_\bullet - \sigma$ relation is based on recent spatially resolved estimates of SMBH masses, and its best-fit slope is between the slopes found by \cite{KO13.1} and \cite{MC13.1} for their best-fit $M_\bullet - \sigma$ relations. Neither observations nor simulations find strong evidence for evolution of the $M_\bullet - \sigma$ relation with redshift (e.g., \citealt{RO06.3, SH15.4}), and as a result we do not add redshift evolution here.

\subsection{Binary SMBH Mass Ratio}
\label{differential}

Models that extrapolate from observations of galaxies in close pairs to a binary SMBH population emitting gravitational waves rely on the simple assumption that there is no appreciable mass growth in the SMBHs as they progress through the galaxy merger (e.g., \citealt{SE13.1,IZ24.1}). In other words, they assume that the mass ratio of the SMBHs scales with the mass ratio of the merging galaxies. 

For the scalings that we use here, $M_\bullet \propto \sigma^{a}$ (where $a =5.07 \pm 0.27$; \citealt{DE19.1}; Section~\ref{bhmass}) and $\sigma \propto M_*^{b }$ (where $b  = 0.293 \pm 0.001$; \citealt{ZA16.1}; Section~\ref{gwb}), the SMBH mass ratio is

\begin{equation}
\label{q_equation}
    \frac{M_{\bullet,2}}{M_{\bullet,1}}=\left(\frac{\sigma_2}{\sigma_1}\right)^{a }=\left(\frac{M_{*,2}^{b }}{M_{*,1}^{b }}\right)^{a }=\left(\frac{M_{*,2}}{M_{*,1}}\right)^{ab} \;,
\end{equation}
which simplifies to $q_\bullet = q_{gal}^{ab}$, or $q_\bullet = q_{gal}^{1.49}$ for the values of $a$ and $b$ that we use here. 

This means that the mass ratio of the SMBHs is less than the mass ratio of the galaxy hosts. The scaling relations that we use imply that, on average, galaxy mergers with $q_{gal} > 0.4$ are needed to produce major mergers of SMBHs ($q_{\bullet} \ge 0.25$; though we note that scatter in the host galaxy--SMBH mass relation broadens the range of $q_{gal}$ values that can lead to major SMBH mergers). In this scenario, a sample limited to major mergers of galaxies will encompass all of the major mergers of SMBHs that are occurring.

However, here we include a model for SMBH growth during the galaxy merger, which causes the SMBH mass ratio to deviate even further from the galaxy mass ratio. Gas that is driven to the nucleus during a galaxy merger can form a circumbinary disk around the binary SMBHs (e.g., \citealt{AR96.2,AR02.2,MI05.1}), and this gas can accrete onto the SMBHs. Several simulations have shown so-called `differential accretion' or `preferential accretion' onto the secondary SMBH, where the secondary SMBH accretes more material than the primary SMBH (e.g., \citealt{BA02.4,FA14.1}). This is due to the secondary SMBH residing closer to the inner edge of the circumbinary disk and intersecting with more gas. Preferential accretion onto the secondary SMBH would not only lead $q_\bullet$ to diverge from the mass ratio of the merging galaxies, but also drive $q_\bullet$ towards unity. Since more equal mass SMBHs emit stronger gravitational waves, it is important to model the evolution of $q_\bullet$ as the SMBH pair progresses to the $\sim$mpc separation gravitational wave regime.

Given that there are no observational constraints on the evolution of $q_\bullet$, we model its evolution using the results from numerical hydrodynamics calculations of accretion onto SMBHs from a circumbinary disk \citep{DU20.1}. They find that the accretion rates evolve as

\begin{equation}
    \frac{\dot{M}_{\bullet,2}}{\dot{M}_{\bullet,1}}=\frac{1}{0.1+0.9q_\bullet} \; ,
\end{equation}
where $\dot{M}_{\bullet,1}$ and $\dot{M}_{\bullet,2}$ are the accretion rates onto the primary and secondary SMBHs, respectively. \cite{SI20.1} show that this model matches the results from the hydrodynamics simulations of circumbinary disk accretion of \cite{MU20.1} as well.

To add an observational constraint on the amount of SMBH mass growth due to accretion during the merger, we apply the results of \cite{FE24.1}. Their observational samples are a catalog of galaxy pairs in SDSS (selected in the same way as the SDSS pairs used to determine the galaxy merger rate in Section~\ref{mergerrate}) and a catalog of post-coalescence mergers from the Ultraviolet Near Infrared Optical Northern Survey. Using these samples, they integrate star formation rate enhancements in merging galaxies to estimate the amount of new stellar mass created by star formation during a galaxy merger. They find that the total cumulative stellar mass growth during a galaxy merger is $10\%$ for massive galaxies (log ($M_* / M_\odot)  \simgt 10.5$), and can be over $20\%$ for less massive systems \citep{FE24.1}. Using the relationship from Equation \ref{q_equation}, a $10\%$ growth in stellar mass becomes a $15\%$ growth in SMBH mass. Because of the uncertainties in the scaling relations behind Equation \ref{q_equation}, we explore both $10\%$ and $20\%$ SMBH mass growth per merger. This is a conservative first step in exploring the impact of differential accretion, as the actual amount of SMBH mass growth may be higher for lower mass galaxies (as described above) and higher redshift galaxies.

The \cite{FE24.1} results are based on a sample of $z<0.3$ galaxies, whereas our model of SMBH mergers extends to $z=1.5$. Galaxy gas fractions and SMBH accretion rates are known to increase with redshift (e.g., \citealt{TA10.2,YA18.1,ZO24.1}), making the \cite{FE24.1} result of $10-20\%$ cumulative stellar mass growth at $z<0.3$ a lower limit for our $0 < z < 1.5$ galaxy sample. 

In the analyses that follow, we use three different prescriptions for $q_\bullet$ evolution during a galaxy merger: zero $q_\bullet$ evolution, the \cite{DU20.1} model with 10\% cumulative growth in the total SMBH mass, and the \cite{DU20.1} model with 20\% cumulative growth in the total SMBH mass. These prescriptions cover the minimum and maximum observationally-based mass growth estimates at $z<0.3$, as well as the zero mass growth scenario for a baseline comparison. 

\begin{figure}[h]
\begin{center}
\includegraphics[width=\linewidth]{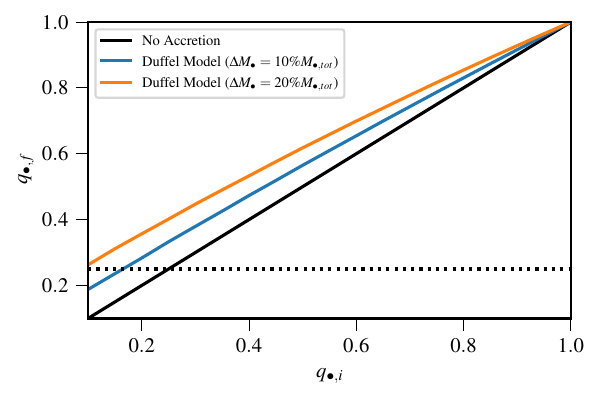}
\end{center}
        \caption{The relationship between the final binary SMBH mass ratio ($q_{\bullet, f}$, the mass ratio at $\sim$mpc separations) and the initial SMBH mass ratio ($q_{\bullet, i}$, the mass ratio at kpc-scale separations) for different models of accretion. Here, $q_{\bullet, i}$ is derived from the mass ratio of the galaxy pair. For no accretion (solid black line), the SMBH mass ratio does not evolve. However, models of preferential accretion onto the secondary SMBH drive $q_{\bullet, f}$ higher, as shown for observationally-based estimates of 10\% (blue solid line) and 20\% (orange solid line) cumulative growth of the total SMBH mass $M_{\bullet, tot}$. The dotted line shows $q_{\bullet, f} = 0.25$, which denotes the cut-off between minor SMBH mergers (below the dotted line) and major SMBH mergers (above the dotted line). We highlight that the differential accretion models bring many SMBH binaries from what would be considered a minor SMBH merger event ($0.1 \leq q_{\bullet} < 0.25$) into the classification for a major SMBH merger event ($0.25 \leq q_{\bullet} \leq 1$). 
        }
        \label{fig1}
\end{figure}

The initial binary SMBH mass ratio $q_{\bullet, i}$, when the SMBHs have kpc-scale separations, is set by populating each pair of merging galaxies with a co-evolving SMBH (Section~\ref{bhmass}). 
We then use the SMBH mass growth prescriptions described above to evolve the binary SMBH mass ratio forward to the final binary SMBH mass ratio $q_{\bullet, f}$, which we define as the mass ratio when the SMBH binaries reach $\sim$mpc separations and their gravitational wave emission enters the pulsar timing array frequency band. Figure~\ref{fig1} shows the evolution from initial to final binary SMBH mass ratio using the three prescriptions of 0, 10\%, and 20\% cumulative SMBH mass growth. We fit polynomials to the two numerical accretion model results and found the following relations
\begin{equation}
\label{qfits}
\begin{split}
    \mathrm{ for} \, \Delta M_\bullet = 10\% &\, M_{\bullet, tot}: \\
     q_{\bullet,f} &= -0.08 q_{\bullet,i}^{2} + 0.99q_{\bullet,i} + 0.09 \\
    \mathrm{ for} \, \Delta M_\bullet = 20\% &\, M_{\bullet, tot}: \\
     q_{\bullet,f} &= -0.13 q_{\bullet,i}^{2} + 0.96q_{\bullet,i} + 0.17 
\end{split}
\end{equation}
where the errors on the coefficients are $\simlt0.001$.

\subsection{SMBH Binary Evolution Timescale}
\label{timescale}

While the merger timescale used in the calculation of the galaxy merger rate in Section~\ref{mergerrate} encodes the time for galaxy pairs to coalesce, the SMBHs in the merging galaxies are still at relatively large, kpc-scale separations at the end of this merger timescale. Therefore, we need to incorporate an additional timescale to account for the SMBH pair evolution at separations $\simlt$kpc. The SMBH binary evolution timescale $T_{evol}$ is the timescale between $\sim$kpc separations and the $\sim$mpc separations when the SMBH binary's gravitational wave emission enters the pulsar timing array band. 

The general processes that drive the SMBH binary's evolution below $\sim$kpc are laid out in \cite{BE80.1}. Dynamical friction drives the evolution of SMBH binaries from $\sim$kpc to $\sim$pc separations, and during this phase the binary experiences a drag force as it passes through a sea of stars and dark matter. Below separations $\sim10$ pc, stellar loss-cone scattering becomes a dominant driver of the binary's evolution. In stellar loss-cone scattering, the SMBH binary intersects the orbit of a star, resulting in a three-body interaction. The star is scattered away, carrying away energy from the SMBH binary and driving the SMBHs closer. The rate at which the loss-cone refills with stars is one of the largest uncertainties in the SMBH binary's evolution (e.g., \citealt{MI03.1,ME13.1,KE17.1}). Below separations $\sim0.1$ pc, a circumbinary gas disk can exert a drag force on the binary as it moves through the gas. Finally, below separations $\sim$mpc, gravitational wave emission shrinks the SMBH binary orbit to coalescence.

Since the specifics of how binaries evolve from kpc to mpc separations are largely unknown, the SMBH binary evolution timescale is also uncertain. Using semi-analytic models applied to the Illustris hydrodynamic cosmological simulations, \cite{SI20.1} form a SMBH binary population and analyze the binary evolution timescales. They define the timescale as the time from when Illustris assumes that each SMBH binary `merged' ($\sim$kpc separations) until coalescence. For the model of differential SMBH mass accretion that we use here (\citealt{DU20.1}; Section~\ref{differential}), \cite{SI20.1} find a probability distribution function of SMBH binary evolution timescales ranging from 0.1 to 10 Gyr, with a median timescale of 1.8 Gyr. They explore six models of SMBH accretion and find similar timescale ranges and median timescales ($1.7-2$ Gyr) for each. Therefore, we explore a range of SMBH binary evolution timescales here, with special attention paid to $T_{evol}=1.8$ Gyr.

We note that the total amount of SMBH mass growth and the binary evolution timescale are independent variables in this work. In reality, these two variables are linked in  galaxy mergers by the specific accretion dynamics present in the circumbinary disk phase. It is beyond the scope of this initial exploration to include the complex interplay of these two parameters, and we leave that analysis to future studies. 

\section{Results and Discussion}
\label{results}

\subsection{Differential Binary SMBH Mass Growth Increases the Number of Major Mergers of SMBHs} 
\label{major}

The mass ratio of SMBHs in a binary system can be used to classify them as minor mergers ($0.1 \leq q_{\bullet} < 0.25$) or major mergers ($0.25 \leq q_{\bullet} \leq 1$). Minor mergers of SMBHs create weaker gravitational waves, since the dimensionless gravitational wave strain amplitude of a binary SMBH depends on SMBH mass ratio as $h_s \propto \frac{q_\bullet}{(1 + q_\bullet)^2}$ (Equation~\ref{chirp}; Equation~\ref{h}). Consequently, minor mergers of SMBHs do not contribute appreciably to the GWB. 

Most models of the GWB do not include minor galaxy mergers (e.g., \citealt{SE13.1,SI16.1}), which makes sense given our calculation that galaxy mergers with $q_{gal}<0.4$ without additional SMBH growth are not expected to produce major SMBH mergers on average (Section~\ref{differential}). In fact, \cite{SE13.1} found that including minor galaxy mergers with $0.1 < q_{gal} < 0.25$ increased the predicted GWB amplitude by only 0.06 dex. 

\begin{figure}[!t]
\begin{center}
    \includegraphics[width=\linewidth]{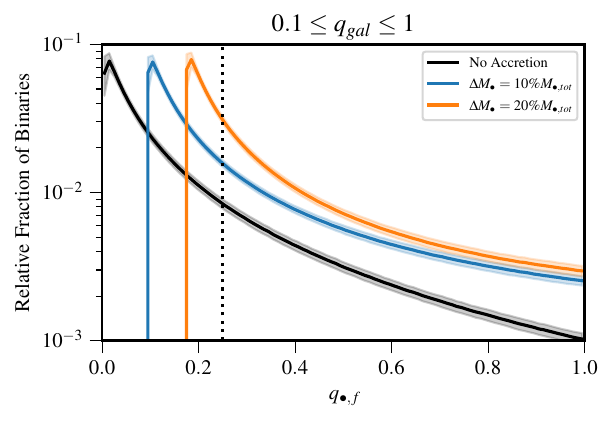}
\end{center}
        \caption{Fraction of SMBH binaries as a function of $q_{\bullet,f}$, the final SMBH mass ratio when the binary enters the pulsar timing gravitational wave band ($\sim$mpc separations), for the population of $0.1 \le q_{gal} \le 1$ galaxy mergers. Solid lines show the median values over 1000 realizations, while the shaded regions show the $1\sigma$ errors. The no accretion model (black line) is where the SMBH mass ratio is derived from the galaxy mass ratio and there is no further SMBH mass growth ($q_{\bullet,f}=q_{\bullet,i}$). The models of preferential accretion onto the secondary SMBH, for observationally-based estimates of 10\% (blue line) and 20\% (orange line) cumulative growth of the total SMBH mass $M_{\bullet,tot}$, drive the fraction of SMBH binaries higher for higher SMBH mass ratios. The dotted vertical line at $q_{\bullet,f}=0.25$ marks the dividing line between minor SMBH mergers (to the left) and major SMBH mergers (to the right). When compared to the no accretion model, the model of 10\% (20\%) total SMBH mass growth increases the fraction of major SMBH mergers from 22\% to 41\% (57\%). 
        }
        \label{fig2}
\end{figure}

However, we find that when we account for differential accretion onto the secondary SMBH, minor {\it galaxy} mergers can create major {\it SMBH} mergers that are significant contributors to the GWB. Figure~\ref{fig2} shows the fraction of binary SMBHs as a function of final SMBH mass ratio for different amounts of cumulative SMBH mass growth during the merger. We find that even modest amounts of differential mass growth (that result in 10\% and 20\% cumulative SMBH mass growth) have an outsized impact, shifting the number of binary SMBHs to higher mass ratios. This is because differential accretion preferentially grows the secondary SMBH's mass, driving increases in the SMBH mass ratio.

Table~\ref{tbl-major} shows the impact of differential accretion on the fraction of galaxy mergers that result in major mergers of SMBHs. 
When compared to the zero accretion model, the preferential accretion model with 10\% cumulative SMBH mass growth in $0.1 \le q_{gal} \le 1$ galaxy mergers increases the fraction of SMBH mergers that are major mergers from 22\% to 41\% (a factor of 1.9 increase). Meanwhile, preferential accretion with 20\% cumulative SMBH mass growth increases the fraction of SMBH mergers that are major mergers from 22\% to 57\% (a factor of 2.6 increase). These results underscore the need to include minor mergers of galaxies in gravitational wave models when incorporating preferential accretion, as minor galaxy mergers may undergo enough mass growth to contribute appreciably to the GWB generated by the population of major SMBH mergers. 

\begin{deluxetable}{llc}
\tabletypesize{\footnotesize}
\tablewidth{0pt}
\tablecolumns{3}
\tablecaption{Fraction of galaxy mergers that produce major SMBH mergers}
\tablehead{
\colhead{$\Delta M_\bullet / M_{\bullet, tot}$} &
\colhead{$q_{gal}$ range} & 
\colhead{Fraction of galaxy mergers that}
\\
\colhead{} & 
\colhead{} & 
\colhead{produce major SMBH mergers} 
}  
\startdata
0\% & $0.1\phantom{0} \le q_{gal} \le 1$ & 0.22 \\ 
0\% & $0.25 \le q_{gal} \le 1$ & 0.35 \\ 
\hline
10\% & $0.1\phantom{0} \le q_{gal} \le 1$ & 0.41 \\ 
10\% & $0.25 \le q_{gal} \le 1$ & 0.64 \\ 
\hline
20\% & $0.1\phantom{0} \le q_{gal} \le 1$ & 0.57 \\ 
20\% & $0.25 \le q_{gal} \le 1$ & 0.80 
\enddata
\tablecomments{Column 1 shows the cumulative amount of differential SMBH mass growth during the merger. Column 2 shows the galaxy mass ratio range of galaxy pairs in the model. Column 3 shows the fraction of all galaxy mergers in the $q_{gal}$ range that result in major mergers ($q_\bullet \ge 0.25$) of SMBHs.}
\label{tbl-major}
\end{deluxetable}

\subsection{Predicted Amplitude of the Gravitational Wave Background}
\label{gwb}

We now model the GWB that originates from a cosmological population of binary SMBHs. The dimensionless amplitude of the strain produced by the population of binary SMBHs is (e.g., \citealt{SE13.1,RA15.1})
\begin{equation}
\label{A_SMBH}
A_{\mathrm{yr}}^2= \int \int \int \Phi_{\bullet} \mathcal{R}_{\bullet} \frac{dV_c}{dz} \frac{dz}{dt} \frac{dt}{d (\ln f_{\mathrm{yr}})}  h_s^2 \; dz \, dM_{\bullet,1} \, dq_{\bullet} \; ,
\end{equation}
where $\Phi_\bullet$ is the SMBH mass function, $\mathcal{R}_\bullet$ is the SMBH merger rate, $V_c$ is the co-moving volume shell, $z$ is the redshift when the SMBH binary's gravitational wave emission enters the pulsar timing array band, $h_s$ is the polarization- and sky-averaged strain from one SMBH binary system, $M_{\bullet,1}$ is the mass of the primary (more massive) SMBH, and $q_\bullet$ is the binary SMBH mass ratio. Here the amplitude $A_{\mathrm{yr}}$ is referenced to the Earth-observed gravitational wave frequency of an inverse year $f_{\mathrm{yr}}=1$ yr$^{-1}$ \citep{SE13.1}.

Our aim is to model $A_{\mathrm{yr}}$ based on observational inputs wherever they are available, and the observations that we use are of galaxies and not of the SMBHs themselves. Therefore, we use a framework of galaxy observations to infer SMBH properties. We describe how we infer each SMBH property from observations of galaxy properties below.

As was done in \citet{SI23.1}, we frame $A_{\mathrm{yr}}$ in terms of galaxy stellar velocity dispersion. To infer the SMBH mass function $\Phi_\bullet (\sigma, z)$, we use the velocity dispersion function described in Section~\ref{bhmass}. To infer the SMBH merger rate $\mathcal{R}_{\bullet}$, we start with the observationally-based galaxy merger rate measured in Section~\ref{mergerrate} and convert the galaxy stellar mass to velocity dispersion via a scaling relation derived from SDSS galaxies \citep{ZA16.1}. We also use this scaling relation to convert the merging galaxy mass ratio $q_{gal}$ (Section~\ref{mergerrate}) into a ratio of velocity dispersions.

We next use the $M_\bullet - \sigma$ relation from \cite{DE19.1} (Section~\ref{bhmass}) to translate velocity dispersion into SMBH mass, which also infers the initial binary SMBH mass ratio $q_{\bullet, i}$. Then we determine the final binary SMBH mass ratio $q_{\bullet, f}$ via the relation that we found in Section~\ref{differential} (Equation~\ref{qfits}), and set $q_\bullet = q_{\bullet, f}$. We carry out this calculation for a range of total SMBH mass growth values ($0$, 10\%, and 20\%; Section~\ref{differential}) and present the range of results here.

To infer the redshift $z$ when the SMBH binary's gravitational wave emission enters the pulsar timing array band, we add the SMBH binary evolution timescale $T_{evol}$ to the redshift of the galaxy merger. $T_{evol}$ is the timescale between the $\sim$kpc SMBH separations when the galaxies coalesce and the $\sim$mpc SMBH separations when the SMBH binary's gravitational wave emission enters the pulsar timing array band. As described in Section~\ref{timescale}, we consider a range of SMBH binary evolution timescales, with particular focus on $T_{evol}=1.8$ Gyr.

Finally, the polarization- and sky-averaged strain contribution from each SMBH binary is (e.g., \citealt{TH87.1})
\begin{equation}
\label{h}
    h_s = \sqrt{\frac{32}{5}} \left(\frac{G \mathcal{M}}{c^3}\right)^{5/3} (\pi f_r)^{2/3} \frac{c}{D_c} \; ,
\end{equation} 
where $\mathcal{M}$ is the chirp mass of the SMBH binary (Equation~\ref{chirp}), $f_r$ is the frequency of the gravitational waves emitted in the rest frame of the SMBH binary (where the Earth-observed gravitational wave frequency is $f=f_r /(1+z)$), and $D_c$ is the proper (comoving) distance to the SMBH binary. As Equation~\ref{chirp} shows, the chirp mass depends on the binary SMBH mass ratio $q_\bullet$, where we set $q_\bullet = q_{\bullet, f}$ as described above, and the total SMBH mass $M_{\bullet, tot}$, which includes the additional SMBH mass growth. 

The limits on the integrals in Equation~\ref{A_SMBH} are set by the bounds on the observational inputs. The redshift range is $0<z<1.5$ (Section~\ref{bhmass}), the velocity dispersion range is $1.85 < \log_{10} (\sigma / \mathrm{km \, s}^{-1}) < 2.6$ \citep{SI23.1}, and the galaxy mass ratio range is $0.1 < q_{gal} < 1$ (Section~\ref{mergerrate}). We then translate the above inputs and their errors into probability distributions for the amplitude $A_{\mathrm{yr}}$ of the GWB, via Equation~\ref{A_SMBH}. This is the first observationally-based calculation of $A_{\mathrm{yr}}$ that incorporates an evolving binary SMBH mass ratio $q_\bullet$ due to differential accretion onto the SMBHs. 

\begin{figure}[!t]
\begin{center}
    \subfigure[]{\includegraphics[width=\linewidth]{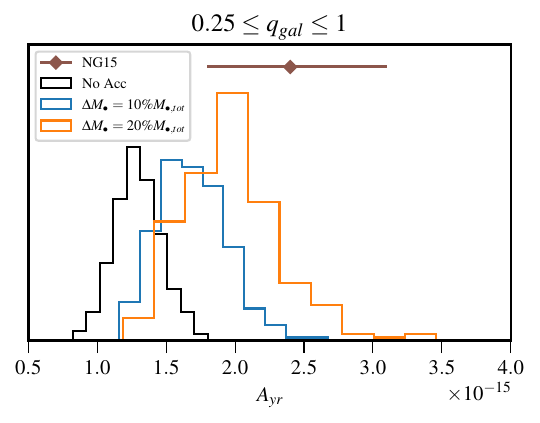}}
    \subfigure[]{\includegraphics[width=\linewidth]{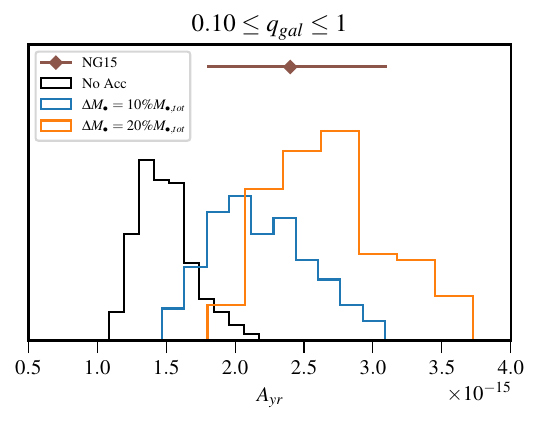}}
\end{center}
        \caption{Probability distribution functions of the predicted GWB amplitude for models of no SMBH accretion (black), differential accretion with 10\% total SMBH mass growth (blue), and differential accretion with 20\% total SMBH mass growth (orange). The SMBH binary evolution timescale is set to 1.8 Gyr, the median timescale found for the differential accretion model we use here \citep{DU20.1,SI20.1}. The brown diamond data point shows the median measured value for the GWB amplitude found in NANOGrav's 15-year data set, with error bars showing the $90\%$ credible region \citep{AG23.1}. The top panel (a) is only for major galaxy mergers ($q_{gal} \geq 0.25$), while the bottom panel (b) is for both major and minor galaxy mergers ($q_{gal} \geq 0.1$). While the no accretion models are similar in the two panels, the other two models change significantly when minor galaxy mergers are included because preferential accretion can turn minor galaxy mergers into major SMBH mergers (see Figure~\ref{fig2}). We note that the model with 10\% total SMBH mass growth due to differential accretion with the inclusion of minor mergers has a median value that most closely matches the median value constrained from pulsar timing array data.
        }
        \label{fig3}
\end{figure}

\begin{figure*}[!t]
\begin{center}
    \includegraphics[width=\textwidth]{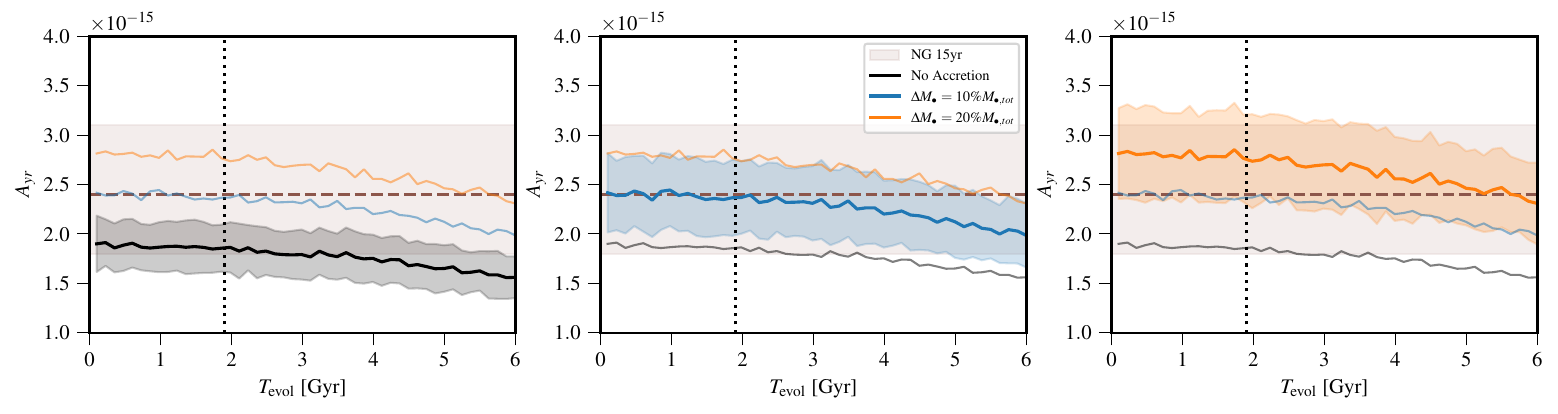}
\end{center}
    \caption{Predicted values of the GWB amplitude for models with no SMBH accretion (black), as well as models of preferential accretion onto the secondary SMBH with total SMBH mass $M_{\bullet,tot}$ growth of 10\% (blue) and 20\% (orange). Each line shows the median values over 1000 realizations, with the shaded regions denoting the $1\sigma$ errors. For clarity we show only one shaded, $1\sigma$ error region per panel: zero (left, gray), 10\% (middle, blue), and 20\% (right, orange) total SMBH mass growth. The SMBH binary evolution timescale $T_{evol}$, which is the timescale for the SMBH binary to evolve from $\sim$kpc separations to $\sim$mpc separations, is allowed to vary. The dotted vertical line illustrates $T_{evol}=1.8$ Gyr, the median timescale found for the model of differential SMBH mass accretion that we use \citep{DU20.1, SI20.1}. The dashed brown horizontal line shows the median amplitude of the GWB found in NANOGrav's 15-year data set (brown shaded area is the $90\%$ credible region; \citealt{AG23.1}). We find that even small amounts (10\%) of differential accretion onto the SMBH binary amplify the amplitude to values consistent with the current constraints from NANOGrav.
    }
    \label{fig4}
\end{figure*}

\subsection{Differential Binary SMBH Mass Growth Strengthens the Gravitational Wave Background}

Figure~\ref{fig3} shows our results for a self-consistent model of the GWB produced by binary SMBHs undergoing differential mass growth with a SMBH binary evolution timescale of $T_{evol}=1.8$ Gyr. We model the differential SMBH mass growth with the analytic expression of \cite{DU20.1}, which \cite{SI20.1} applied to the Illustris cosmological simulations and found that the median timescale was $T_{evol}=1.8$ Gyr (Section~\ref{timescale}). Therefore, although there is much uncertainty surrounding the specifics of SMBH binary evolution from $\sim$kpc to $\sim$mpc separations and the corresponding timescale, $T_{evol}=1.8$ Gyr provides self consistency between the model of differential SMBH mass growth and the corresponding SMBH binary evolution timescale. We find that a model of differential accretion onto the binary SMBHs that increases the total SMBH mass by 10\% brings the predicted GWB amplitude into agreement with the median value measured by NANOGrav in their 15-year data set \citep{AG23.1}.

If we allow the SMBH binary evolution timescale to vary, our results are shown in Figure~\ref{fig4}. Longer timescales lead to a lower rate of SMBH mergers, which decreases the predicted GWB amplitude. For longer SMBH binary evolution timescales, we find that the zero accretion model becomes inconsistent with current measurements of the GWB amplitude, which is consistent with other studies (e.g., \citealt{AG23.2}), whereas models with SMBH accretion are consistent with these observations. 

It is worth noting how gradual the decrease in the GWB amplitude is with increasing $T_{evol}$, which is different from the behavior seen in models that utilize a galaxy stellar mass function (GSMF) rather than a stellar velocity dispersion function (VDF). The GSMF uses $M_\bullet-M_{bulge}$ to obtain SMBH mass and therefore derives SMBH mass from the bulge mass alone, whereas the VDF uses a galaxy's total stellar mass, effective radius, and S{\' e}rsic index to derive SMBH mass. The population of SMBH masses derived from $M_\bullet-M_{bulge}$ shifts to lower SMBH mass as redshift increases, because bulge mass has an inverse relationship with redshift (e.g., \citealt{LE20.1}). However, higher redshift galaxies are more compact with smaller effective radii (e.g., \citealt{VA14.1}) and inferred velocity dispersion increases with decreasing effective radius (e.g., \citealt{BE11.2}), which means that inferred velocity dispersions (and therefore the SMBH masses derived from the velocity dispersions) stay relatively high with increasing redshift \citep{TA22.1}. As a result, the VDF produces a much greater number density of the highest mass SMBHs than the GSMF does \citep{LE20.1,TA22.1}. The difference is most notable at $z>0.5$, where the GSMF produces many fewer high mass SMBHs than the VDF (e.g., \citealt{MA23.1}). The amplitude of the GWB is most impacted by the highest mass SMBHs, and these SMBHs need to have a short enough $T_{evol}$ to be able to coalesce by $z=0$ and contribute to the observable GWB. The GWB amplitude inferred from the VDF is less sensitive to changes in $T_{evol}$ because the VDF produces a large population of high mass SMBHs at $z>0.5$ that can coalesce by $z=0$ for a relatively wide range of $T_{evol}$ values. In contrast, the GWB amplitude inferred from the GSMF produces high mass SMBHs primarily at $z<0.5$ and consequently requires a short $T_{evol}$ for the SMBHs to be able to coalesce by $z=0$ and contribute to the observable GWB (e.g., \citealt{SI16.1,AG23.2}).

Overall, we find that any amount of differential binary SMBH mass growth increases the amplitude of the GWB. This is because differential SMBH mass growth preferentially increases the secondary SMBH's mass, which increases the binary SMBH mass ratio. A binary SMBH system with a larger mass ratio has a larger chirp mass (Equation~\ref{chirp}), and a binary system with a larger chirp mass creates a larger gravitational wave strain (Equation~\ref{h}). Since the GWB is composed of the aggregate strains from a cosmological population of SMBH binaries, an increase in strain also increases the amplitude of the background. We find that even small amounts of differential binary SMBH mass growth are sufficient to bring models of the GWB in line with the observational measurements.

\subsection{Implications of Differential Accretion for Continuous Wave, LISA, JWST, and LIGO Sources}

Our analysis thus far has focused on differential accretion onto SMBHs and its effect on the GWB that is observed by pulsar timing arrays. However, differential accretion can also impact different types of black holes, including those observed by different facilities. The model that we use for differential accretion onto a black hole pair (\citealt{DU20.1}; Section~\ref{differential}) is scale free and can be applied to different masses of black holes, but the mass growth that we assume (10\% and 20\% of the total black hole mass; Section~\ref{differential}) is tailored to SMBHs in $z<0.3$ galaxy mergers and does not necessarily apply to other scenarios. As one example of the broader effects of differential accretion, it is known to result in more misalignment in the spins of merging black holes, which in turn produces higher gravitational wave recoil velocities (e.g., \citealt{GE15.1}).  Here we consider the effect of differential accretion on continuous wave sources, as well as on the black holes detected by the Laser Interferometer Space Antenna (LISA; \citealt{AM17.1}), James Webb Space Telescope (JWST), and the Laser Interferometer Gravitational-wave Observatory (LIGO).

\subsubsection{Shorter Time to the First Continuous Wave Source Detection}

Pulsar timing arrays are also capable of detecting individual SMBH binaries whose gravitational wave signals are strong enough to stand out against the GWB (e.g., \citealt{SE09.1,RO15.2}). These continuous wave sources are so named because they are expected to emit almost monochromatic gravitational waves continuously over the decades that they are detectable within the pulsar timing array frequency band (e.g., \citealt{CO10.2}). The most detectable continuous wave sources are those with the highest SMBH masses ($\sim10^9 - 10^{10}$ $M_\odot$; e.g., \citealt{GA24.1}). Pulsar timing arrays have not yet detected a continuous wave source \citep{AG23.3,EP24.2}, but it is the next hotly anticipated event to come from pulsar timing array observations.

Current models of continuous wave sources do not account for differential accretion onto the SMBH binary, and predict that $\sim$20 year total observing baselines are required for pulsar timing arrays to resolve continuous wave sources above the GWB (e.g., \citealt{RA15.1,KE18.1,GA24.1}). However, preferential accretion onto the secondary SMBH would strengthen the gravitational wave signals from continuous wave sources, leading to a shorter required observing baseline and a faster detection of the first continuous wave source in pulsar timing array data.

\subsubsection{Stronger Gravitational Waves from Massive Black Hole Binaries Observed by LISA}

Preferential accretion onto the secondary black hole would strengthen the gravitational wave signals from black hole binaries (expected total masses $\sim10^3$ to $10^7$ $M_\odot$) detected by LISA. LISA can detect gravitational waves from binary black holes out to high redshifts (up to and beyond $z=20$), when galaxies were more gas-rich and black hole accretion rates were higher (e.g., \citealt{TA13.1,MA23.2}). The cumulative mass growth of these higher redshift black holes may therefore be larger as well. Since LISA is sensitive to higher redshift black hole binaries than pulsar timing arrays, differential accretion may cause an even larger enhancement of the gravitational wave signal for LISA than it does for pulsar timing arrays.

\subsubsection{Faster Assembly of High Redshift Black Holes Observed by JWST}

JWST has revealed a population of surprisingly massive black holes at high redshifts (e.g., \citealt{PA23.1, MA24.1}), and differential accretion could help explain how these black holes built up their masses so quickly. 
Mergers dominate early black hole growth (e.g., \citealt{BH25.1}) and these mergers could trigger accretion onto the black holes (e.g., \citealt{TR24.1}). If this accretion follows the form of preferential accretion onto the secondary black hole, then both the merging black hole mass ratio and the chirp mass $\mathcal{M}$ would increase and the gravitational wave inspiral timescale ($\tau_{GW} \propto \mathcal{M}^{-5/3}$) would decrease. Consequently, differential accretion would propel faster buildup of the black hole mass via mergers, helping the massive black holes to be in place by the redshifts at which they are being observed by JWST.

\subsubsection{Stronger Gravitational Waves from Stellar Mass Black Hole Binaries Observed by LIGO}

LIGO and Virgo have detected gravitational waves from binary black holes with masses that reside in the pair instability mass gap (e.g., \citealt{AB20.1,NI23.1}). The mass gap, which ranges from 50 to 130 $M_\odot$, arises because stellar cores of this mass are expected to undergo pair-instability supernovae, thus preventing the formation of black holes in this mass range (e.g., \citealt{WO21.1}). Mergers of smaller black holes, with masses that are permitted by stellar evolution models, have been proposed as a possible pathway to build up the more massive black hole binaries detected by LIGO and Virgo.

The stellar progenitors of these binary black holes lose mass through several stages of their evolution, and this mass could form a circumbinary disk encompassing the binary. In particular, in two isolated binary formation scenarios (a classical common-envelope scenario, where there are several stages of mass transfer; and a chemically homogeneous evolution scenario, where two tidally distorted binary stars remain compact and experience strong internal mixing) a fraction of mass is shed that can be retained to form a circumbinary disk (e.g., \citealt{DE17.2} and references therein). The isolated binary formation scenarios can favor the formation of unequal mass black hole binaries (e.g., \citealt{OL24.1}), but the binary mass ratio can change due to accretion of material from the circumbinary disk. As the binary dynamically evolves through the disk phase, there can be preferential accretion onto the secondary black hole, following the prescription we use here (\citealt{DU20.1}; Section~\ref{differential}), which increases the binary mass ratio and chirp mass and consequently creates stronger gravitational waves. Therefore, ground-based gravitational wave detectors such as LIGO-Virgo-KAGRA may be biased towards detecting the population of binaries that have circumbinary disks and experience preferential accretion onto the secondary source. Preferential accretion could explain why the black hole binaries found by LIGO and Virgo are preferentially more equal mass binaries \citep{FI20.1}.

\section{Conclusions}
\label{conclusions}

Galaxy mergers create binary SMBHs and also drive central inflows of gas that can accrete onto the binaries. Both observations and simulations suggest that the secondary, less massive SMBH accretes more material than the primary, more massive SMBH in a SMBH pair. For the first time, we carry out an observationally-based analysis of the effect of this preferential accretion on the amplitude of the GWB produced by the cosmological population of SMBH binaries. 

Our predictions for the GWB amplitude are based on observational inputs whenever they are available. We build our model based on observations of the velocity dispersion function; SDSS spectroscopic close pairs of galaxies; the observationally-based $M_\bullet - \sigma$ relation for SMBH masses; a simulations-based model for preferential accretion onto the secondary SMBH as the merger progresses; and scenarios of 0\%, 10\%, and 20\% cumulative growth in the total SMBH mass, which are motivated by observational measurements of the total stellar mass growth during mergers. The SMBH binary evolution timescale, which is the timescale for the binary to dynamically evolve from $\sim$kpc to $\sim$mpc separations, is observationally unconstrained. Consequently, we allow it to vary while also paying special attention to the 1.8 Gyr median timescale predicted for our model of differential SMBH mass growth. 

Our main results are summarized below.

1. Preferential accretion onto the secondary SMBH drives the SMBH mass ratio upwards, leading to an increase in the number of major mergers of SMBHs (Figure~\ref{fig1}). In this way, {\it minor} mergers of galaxies can create {\it major} mergers of SMBHs. Preferential accretion with 10\% (20\%) cumulative SMBH mass growth leads to a factor of 2 (3) increase in the fraction of SMBH mergers that are major mergers (Table~\ref{tbl-major}; Figure~\ref{fig2}).

2. Differential accretion with a 10\% increase in total SMBH mass brings the predicted amplitude of the GWB into agreement with recent measurements from NANOGrav (Figure~\ref{fig3}). This predicted amplitude includes an SMBH binary evolution timescale of 1.8 Gyr, which is the median timescale found in cosmological simulations that apply the model of differential accretion that we use here.

3. If we allow the SMBH binary evolution timescale to be a free parameter in our model of the GWB, then the predicted amplitude decreases with increasing timescale. However, differential accretion onto the binary SMBHs can strengthen the predicted amplitude into agreement with the NANOGrav observational measurements (Figure~\ref{fig4}).

4. Differential accretion strengthens the gravitational wave signals from any binary black hole system embedded in a circumbinary disk, including those observed by ground-based gravitational wave detectors (e.g., LIGO-Virgo-KAGRA) and by LISA. These stronger signals would lead to a shorter time to the first detection of an individual SMBH binary emitting continuous waves in pulsar timing array data. Differential accretion can also speed up the formation of surprisingly massive black holes seen at high redshifts by JWST.

In summary, we find that just one addition to models of the cosmic binary SMBH population -- preferential accretion onto the secondary SMBH -- is all that is needed to bring the predicted GWB into alignment with pulsar timing array observations. We did not make any adjustments to the local number density of SMBHs or the SMBH mass function in our models.

So far, the models of differential SMBH accretion are based on simulations. Future observations of the evolution of SMBH mass ratios as mergers progress could provide additional insight into the role of differential accretion in the evolution of binary SMBHs and their resultant gravitational waves.

\acknowledgements We thank Laura Blecha, Kayhan G{\"u}ltekin, Magda Siwek, and Sarah Vigeland for useful discussions. J.M.C. acknowledges support from NSF AST-1847938 and NSF AST-2510894. J.S. is supported by NSF Physics Frontiers Center award 2020265 and received previous support from NSF AST-2202388.

\bibliographystyle{aasjournalv7}

\end{document}